\documentclass[%
 reprint,
 superscriptaddress,
 amsmath,amssymb,
 aps,
]{revtex4-2}

\usepackage{graphicx}
\usepackage{dcolumn}
\usepackage{bm}
\usepackage{xcolor}
\definecolor{teal}{rgb}{0,0,0}

\begin{document}

\preprint{APS/123-QED}

\title{Birefringence-mediated enhancement of the magneto-optical activity in anisotropic magnetic crystals}

\author{D.O. Ignatyeva}
\email[]{ignatyeva@physics.msu.ru}
\affiliation{Faculty of Physics, M.V. Lomonosov Moscow State University, 119991 Moscow, Russia}
\affiliation{Russian Quantum Center, 121205 Moscow, Russia}
\affiliation{Institute of Physics and Technology, V.I. Vernadsky Crimean Federal University, 295007 Simferopol, Crimea}

\author{A.A. Voronov}
\thanks{Current address of work: Faculty of Physics, University of Vienna, 1090 Vienna, Austria}
\affiliation{Faculty of Physics, M.V. Lomonosov Moscow State University, 119991 Moscow, Russia}
\affiliation{Russian Quantum Center, 121205 Moscow, Russia}

\author{P.V. Shilina}
\affiliation{Russian Quantum Center, 121205 Moscow, Russia}
\affiliation{Faculty of Physics, HSE University, 101000 Moscow, Russia}

\author{P.O. Kapralov}
\affiliation{Russian Quantum Center, 121205 Moscow, Russia}

\author{S.V. Yagupov}
\affiliation{Institute of Physics and Technology, V.I. Vernadsky Crimean Federal University, 295007 Simferopol, Crimea}

\author{Y.A.~Mogilenec}
\affiliation{Institute of Physics and Technology, V.I. Vernadsky Crimean Federal University, 295007 Simferopol, Crimea}

\author{M.B. Strugatsky}
\affiliation{Institute of Physics and Technology, V.I. Vernadsky Crimean Federal University, 295007 Simferopol, Crimea}

\author{V.I. Belotelov}
\affiliation{Faculty of Physics, M.V. Lomonosov Moscow State University, 119991 Moscow, Russia}
\affiliation{Russian Quantum Center, 121205 Moscow, Russia}
\affiliation{Institute of Physics and Technology, V.I. Vernadsky Crimean Federal University, 295007 Simferopol, Crimea}

\date{\today}

\begin{abstract}
\textcolor{teal}{Optical anisotropy is usually treated as an unfavorable condition for the magneto-optical measurements since it is known to diminish the Faraday rotation concerning the case of the isotropic medium. Here we show that the situation could be quite opposite: a phenomenon of birefringence mediated enhancement of the magneto-optical activity appears if the incident light polarization and angle of incidence are set properly. The present study relies on the experimental, analytical, and numerical studies of iron borate $\mathrm{FeBO_3}$ crystals.  We demonstrate a significant increase of the magneto-optical activity resulting in nearly $100\%$ magneto-optical light modulation magnitude. The approach applies to other types of birefringent crystals with the magneto-optical response that makes it crucial for various practical applications, including magneto-optical microscopy, pump-probe studies, and others.}
\end{abstract}

\maketitle


Light interaction with the magnetic crystals strongly depends on the optical and magnetic properties they possess. \textcolor{teal}{Faraday effect is the well-known phenomenon of rotation of light polarization when light propagates through a magnetized medium. The Faraday rotation is accumulated along the whole pass of light in the isotropic material. Therefore, by increasing a sample thickness one may obtain as strong rotation as desired. The situation drastically changes if the material is optically anisotropic since in this case the birefringence also affects the light polarization. This is the reason why for many years, the Faraday effect in the birefringent crystals has been studied in configurations when light either propagates along the optical axis or has a pure ordinary (or extraordinary) initial polarization. A requirement to send the light parallel to the optical axis significantly limits the experimental capabilities. Moreover, in some cases, such configuration cannot be implemented, for example, if magnetization is perpendicular to the optical axis, or if the optical axis is canted at a large angle to its face. The presence of optical birefringence, in its turn, significantly influences and limits the value~\cite{andlauer1976optical,liu2014competition,de2017effect} and even changes the sign of the Faraday effect~\cite{kurtzig1971faraday,donovan1962theory,chetkin1971magnetooptical}. From 1936 to nowadays, plenty of magneto-optical studies showed 'the decrease in the magnetic rotation with birefringence'~\cite{ramaseshan1951faraday} (see also a historical overview~\cite{suppl}) and focused on the ways to eliminate optical anisotropy. This problem has become especially acute nowadays, with the extensive studies of the antiferromagnetic materials most of which possess birefringence.} 

Antiferromagnetic crystals attract much attention since their magnetization state can be controlled at THz frequencies~\cite{nvemec2018antiferromagnetic,bossini2017femtosecond}. At the same time, most antiferromagnetic materials are characterized by high optical and magnetic anisotropy. \textcolor{teal}{Thus, a birefringence-induced decrease of the Faraday rotation limits the static and dynamic magneto-optical studies of these materials and causes significant obstacles for their applications.} Bearing in mind the importance of \textcolor{teal}{the magneto-optical effects for the domain vizualization~\cite{scott1974magnetic,cheong2020seeing}}, probing of the ultrafast magnetic dynamics~\cite{mashkovich2019terahertz}, sensing~\cite{zubov2017effect}, biosensing~\cite{borovkova2020high,ignatyeva2016high} and magnetometry~\cite{ignatyeva2021vector}, we focus our current study on the peculiarities of the Faraday effect that arise from the optical anisotropy of an antiferromagnetic crystal.  

\textcolor{teal}{To illustrate the proposed approach, we consider} an iron borate $\mathrm{FeBO_3}$ crystal \textcolor{teal}{that has} a quasi-antiferromagnetic mode \textcolor{teal}{of} 0.5~THz frequency~\cite{mashkovich2019terahertz, mikhaylovskiy2020resonant, afanasiev2014laser,kalashnikova2008impulsive,kalashnikova2015ultrafast}. Usually, $\mathrm{FeBO_3}$ crystal plates are cut parallel to (0,0,1) crystal plane, and their magnetization vector lies in the sample plane. \textcolor{teal}{The in-plane magnetization produces zero magneto-optical Faraday effect for normal light incidence~\cite{kalashnikova2008impulsive}}. On the other hand, $\mathrm{FeBO_3}$ is a uniaxial crystal with an optical axis of [0,0,1] direction~\cite{markovin2008optical}. Thus, \textcolor{teal}{detuning from the normal incidence results in the appearance of strong birefringence, which, according to the previously conducted studies~\cite{suppl}, limits the Faraday effect value. In contrast to the recent studies that exploit the features of the nanoplasmonic~\cite{lodewijks2014magnetoplasmonic,armelles2013magnetoplasmonics,ignatyeva2012surface,pomozov2020two,bossini2016magnetoplasmonics} and nanophotonic~\cite{ignatyeva2020all,voronov2020magneto,chernov2020all} structures, we show that optical anisotropy of the crystal itself provides a powerful tool for the drastic increase of the magneto-optical effects if the parameters of the light (its polarization and angle of incidence) are set properly.}

\textcolor{teal}{The presented approach is applicable to any types of the birefringent magneto-optical materials, including a wide range of canted antiferromagnets, such as $\mathrm{FeBO_3}$~\cite{mashkovich2019terahertz}, $\mathrm{TuFeO_3}$ ~\cite{baierl2016nonlinear, grishunin2018terahertz}, $\mathrm{EuFeO_3}$~\cite{mikhaylovskiy2014terahertz,watanabe2017observation}, $\mathrm{DyFeO_3}$~\cite{afanasiev2016control}, $\mathrm{BiFeO_3}$~\cite{khan2020ultrafast}, and various heterostructures, $\mathrm{Fe/RFeO_3}$ (R=Dy or Er)~\cite{tang2018ultrafast} to name just a few.} 

\begin{figure}[htbp]
\centering
(a)\\
\includegraphics[width=0.5\linewidth]{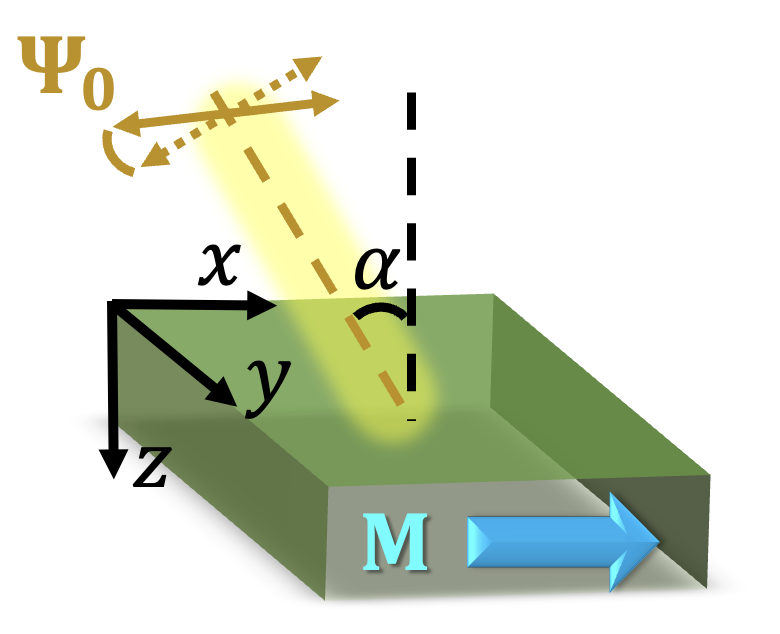} \\
(b)~~~~~~~~~~~~~~~~~~~~~~~~~~~~~~~~~~(c) \\
\includegraphics[width=0.49\linewidth]{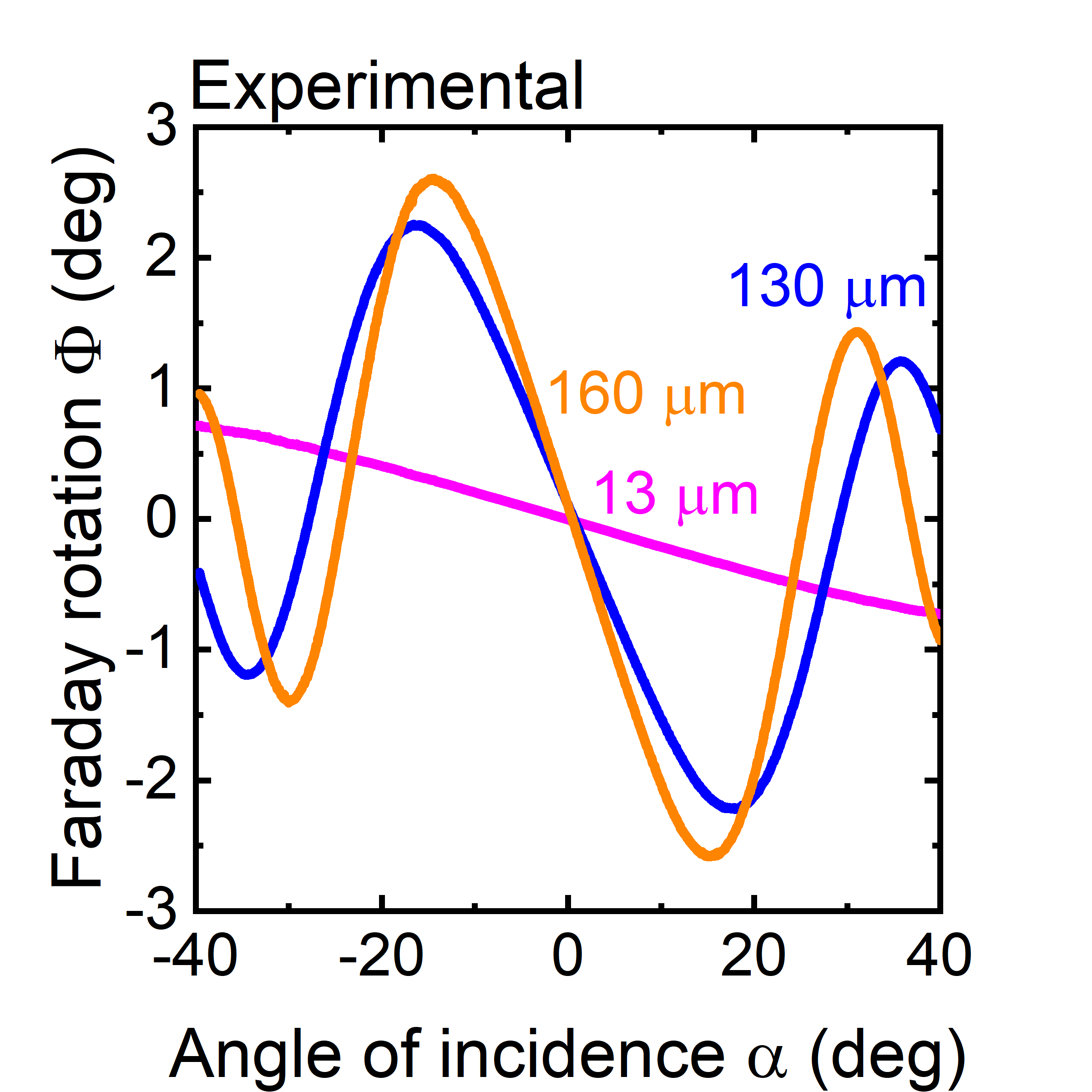}
\includegraphics[width=0.49\linewidth]{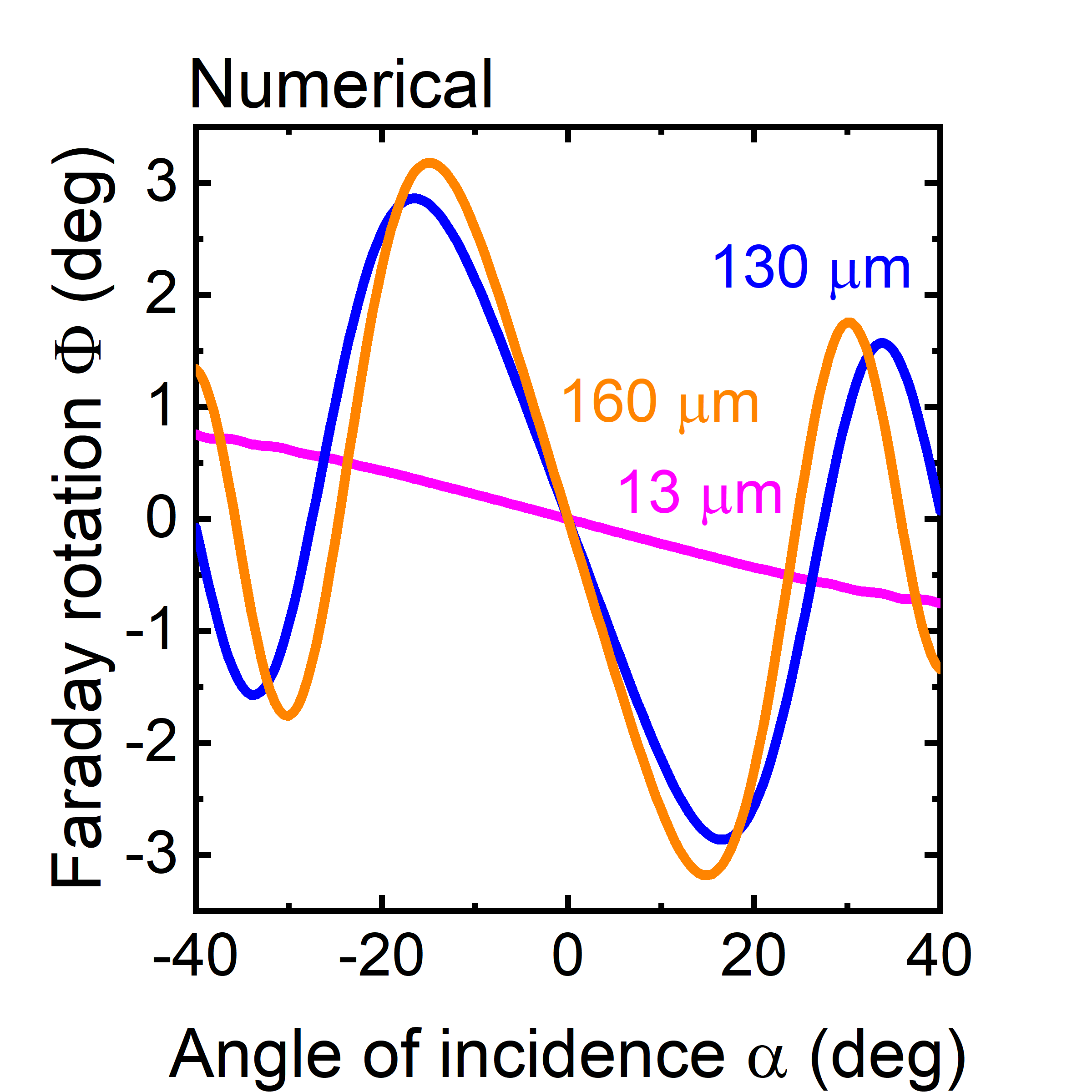}
\caption{\textcolor{teal}{Faraday rotation of $\mathrm{FeBO_3}$ crystal. (a) Schematic representation of the considered configuration. (b,c) Angular spectra of the Faraday rotation (initial polarization is $\Psi_0=0^\circ$, wavelength is $\lambda=0.52 \mu m$) obtained (b) experimentally and (c)} numerically for different thicknesses of $\mathrm{FeBO_3}$ crystals (see legends).}
\label{fig: MCD and Far}
\end{figure}

First, we derive the analytical formulas describing the magneto-optical effect in the $\mathrm{FeBO_3}$ crystal plate taking into account its optical and magnetic anisotropy. Due to the Dzyaloshinskii-Moriya exchange interaction spins of Fe ions are slightly canted and net magnetization appears.  \textcolor{teal}{Let's assume the crystal plate cut perpendicular to [0,0,1] and thus having optical axis along its normal (Z-axis) and magnetization vector $\mathbf{M}$ in-plane (XY-plane)} (Fig.~\ref{fig: MCD and Far}a). Considering $\mathbf{M}$ to be in the plane of incidence (XZ-plane), i.e. along X-axis, one may obtain the elements of permittivity tensor $\hat{\varepsilon}$: $\varepsilon_{xx}=\varepsilon_{yy}=\varepsilon_{o}$, $\varepsilon_{zz}= \varepsilon_{e}$, $\varepsilon_{zy}=-\varepsilon_{yz}= ig$, where $g$ is the gyration coefficient.

To find the refractive indices and polarization vectors of the eigenmodes propagating in the birefringent magneto-optical crystal, one may use the wave equation:
$\mathbf{n}^2\mathbf{E}-\mathbf{n}(\mathbf{n,E})=\hat{\varepsilon}\mathbf{E}$,
where $\mathbf{n}$ is the refraction vector, and $\mathbf{E}$ is the electric field of light, and obtain the following eigenmode equation:
\begin{equation}
\begin{pmatrix}
n_z^2-\varepsilon_o & 0 & -n_z^2 \sin{\alpha} \\
0 & n_z^2-\varepsilon_o\cos^2{\alpha} & ig \\
-n_z^2 \sin{\alpha} & -ig & \varepsilon_o\sin^2{\alpha}-\varepsilon_e
\end{pmatrix}
\begin{pmatrix}
E_x \\
E_y\\
E_z
\end{pmatrix}=0
\label{matrix_equation}
\end{equation}
for the light propagating inside the $\mathrm{FeBO_3}$ crystal at an angle $\alpha$ with respect to the optical axis of the crystal ($z$-axis). \textcolor{teal}{General solution of Eq.~\eqref{matrix_equation} is very bulky~\cite{donovan1962theory} which makes its qualitative analysis nearly impossible. Therefore,in the follows we make several assumptions that would help us to obtain rigorous analytical formulas for the general case of the arbitrary angle of incidence and polarization state, in contrast to the other studies considering only a certain configuration~\cite{ramaseshan1951faraday,kurtzig1971faraday,chetkin1971magnetooptical,liu2014competition,de2017effect,nakagawa2019determination}}.

Typically, the gyration coefficient $g\sim10^{-3}$ is small compared both to the permittivity $g\ll \varepsilon_{ii}$ and birefringence: $g\ll |\varepsilon_o-\varepsilon_e|$, so that linear in $g$ approximation could be used, and the quadratic and higher terms are neglected to simplify the problem. We will also use a linear in $\Delta \varepsilon=\varepsilon_o-\varepsilon_e$ approximation $\Delta \varepsilon \ll \varepsilon_{ii}$ and take into account the accessible range of the light incidence angles $\alpha$ inside $\mathrm{FeBO_3}$ crystal $-\arcsin{\varepsilon_o^{-1/2}}\le \alpha \le \arcsin{\varepsilon_o^{-1/2}}$. Since the inequality $g \ll \Delta \varepsilon \sin \alpha$ is valid for the wide angular range except for a very small region near the normal incidence $|\alpha|<g/\Delta \varepsilon\sim 0.01$, it will also be used to simplify the analysis.

The rigorous solution of Eq.(~\ref{matrix_equation}) with the assumptions listed above gives two eigenmodes with the refractive indices $n_{z+}^2=\varepsilon_o\cos^2{\alpha}$ and $n_{z-}^2=\varepsilon_o\cos^2{\alpha}-\Delta\varepsilon\sin^2{\alpha}$ corresponding to the refractive indices of the ordinary ($n_{z+}$) and extraordinary ($n_{z-}$) waves in the non-gyrotropic crystals. Eigen polarizations of these modes:
\begin{equation}
\mathbf{E}_{+}=
\begin{pmatrix}
-i\frac{g }{\Delta \varepsilon \sin{\alpha}}\cos\alpha \\
1\\
i\frac{g }{\Delta \varepsilon}
\end{pmatrix},
\label{E_eigenmodes_1}
\end{equation}
\begin{equation}
\mathbf{E}_{-}=
\begin{pmatrix}
1 \\
-i\frac{g }{\Delta \varepsilon \sin{\alpha}}\frac{1}{\cos\alpha}\\
-\tan{\alpha} (1 + \frac{\Delta \varepsilon}{2\varepsilon_o }(2+\tan^2 \alpha))
\end{pmatrix},
\label{E_eigenmodes_2}
\end{equation}
acquire the magneto-optically induced ellipticity and tend to the linearly polarized ordinary and extraordinary waves if $g\rightarrow0$. 

There is a qualitative difference in the origin of the magneto-optical effects in the cases of the anisotropic and isotropic materials. In isotropic or quasi-isotropic ($g\gg \Delta \varepsilon$) magnetic crystals, the magnetic field lifts the degeneracy between the propagation constants of two circularly polarized eigenmodes, but the eigenmode field doesn't depend on the magnetic field. On the contrary, in the case of the anisotropic crystals ($g\ll \Delta \varepsilon$), magnetization changes the eigen polarization of the linear modes to elliptical without any impact on the propagation constants of these modes.

The electromagnetic field of the light transmitted through the birefringent magnetic crystal could be calculated $\mathbf{E}^{out}=\hat{J}\mathbf{E}^{in}$ using Jones matrix $\hat{J}$:
\begin{equation}
\hat{J}=
\begin{pmatrix}
\cos{\phi} - i\sin{\phi} & 2 \frac{g }{\Delta \varepsilon \sin{\alpha}}\cos\alpha \sin\phi \\
-2\frac{g }{\Delta \varepsilon \sin{\alpha}}\frac{1}{\cos\alpha}\sin\phi & \cos{\phi} + i\sin{\phi},
\end{pmatrix},
\label{Jones_matrix}
\end{equation}
where $\phi=\Delta \varepsilon k_0 d \tan \alpha \sin {\alpha}  / (4\sqrt{\varepsilon_o})$ is the birefringent phase shift acquired in the crystal of thickness $d$. The polarization modification could be described in terms of \textcolor{teal}{complex} polarization angle determined as the ratio of s- and p- polarized complex amplitudes: $\Psi=\arctan({E_s/E_p})$\textcolor{teal}{~\cite{czyz2007complex}. In this case, the real part of $\Psi$ determines the tilt of light $\mathbf{E}$-vector to the incidence plane (see Fig.\ref{fig: MCD and Far}a), and its imaginary part describes the light ellipticity. For example,} $\Psi=0^\circ$ and $\Psi=90^\circ$ correspond to pure p- and s-polarizations, and $\Psi=\pm i90^\circ$ describe the left- and right- circular polarizations of light. \textcolor{teal}{Polarization modification inside a birefringent magnetic crystal has purely optical and magneto-optical contributions. Analysis of the Jones matrix (Eq.~\ref{Jones_matrix}) shows that the magneto-optical variation of $\Re(\Psi)$ corresponding to the Faraday rotation angle $\Phi$ is}:
\begin{equation}
    \Phi=-\frac{g}{\Delta \varepsilon \sin \alpha} \frac{\sin{2\phi} ( 1-2\sin^2\phi \sin^2 2\Psi_0)}{\cos^2 2\phi+\sin^2 2\phi \cos^2 2\Psi_0},
    \label{Far_rotation}
\end{equation}
in the case of the arbitrary \textcolor{teal}{initial} polarization angle $\Psi_0$ of the linearly polarized light. A significant dependence of $\Phi$ on the angle of incidence and crystal thickness causes the \textcolor{teal}{peculiar} oscillatory patterns of the angular dependencies of the observed magneto-optical effects.

An experimental study was carried out for $\mathrm{FeBO_3}$ crystals of different thicknesses (see Fig. S2 in Supplementary materials~\cite{suppl}). The crystals were grown by the modified solution-in-melt method~\cite{yagupov2018development}. \textcolor{teal}{The measurements were performed at the wavelength $\lambda=0.52\mu m$ corresponding to the one of the transparency windows of $\mathrm{FeBO_3}$ crystal in the visible range where the absorption coefficient is $\alpha\simeq40$~cm$^{-1}$, and the magneto-optical figure of merit is $14^\circ/$dB~\cite{kurtzig1970magneto} close to the iron garnet one~\cite{onbasli2016optical}. We have also performed the numerical simulations of the magneto-optical response that take into account Fresnel reflection at the crystal interfaces}. 

\textcolor{teal}{First, the Faraday rotation was studied in a well-known configuration of light polarization being in the plane of incidence, which corresponds to the extraordinary wave ($\Psi_0=0^\circ$), and thus purely optical polarization modification was excluded}. Figure~\ref{fig: MCD and Far}b shows that the results of the experimental measurements are in good agreement with the numerical simulations \textcolor{teal}{and previous studies (e.g.~\cite{kurtzig1971faraday})}. \textcolor{teal}{In the fixed angular range $\theta>0$ (or $\theta<0$) the angular dependence of the Faraday rotation is smooth for the thin crystals. The thick crystals, where the birefringence phase shift exceeds $\pi/2$, exhibit a sign-changing oscillatory pattern of \textcolor{teal}{$\Phi\propto\sin 2\phi$} according to Eq.~\eqref{Far_rotation}. Numerical simulations show that, indeed, the birefringence  diminishes the Faraday rotation compared to the case of the isotropic medium with the same specific Faraday rotation angle (see also Fig. S3 in Supplementary materials~\cite{suppl}).}

\begin{figure}[htbp]
\centering
(a) \\
~~~\includegraphics[width=0.83\linewidth]{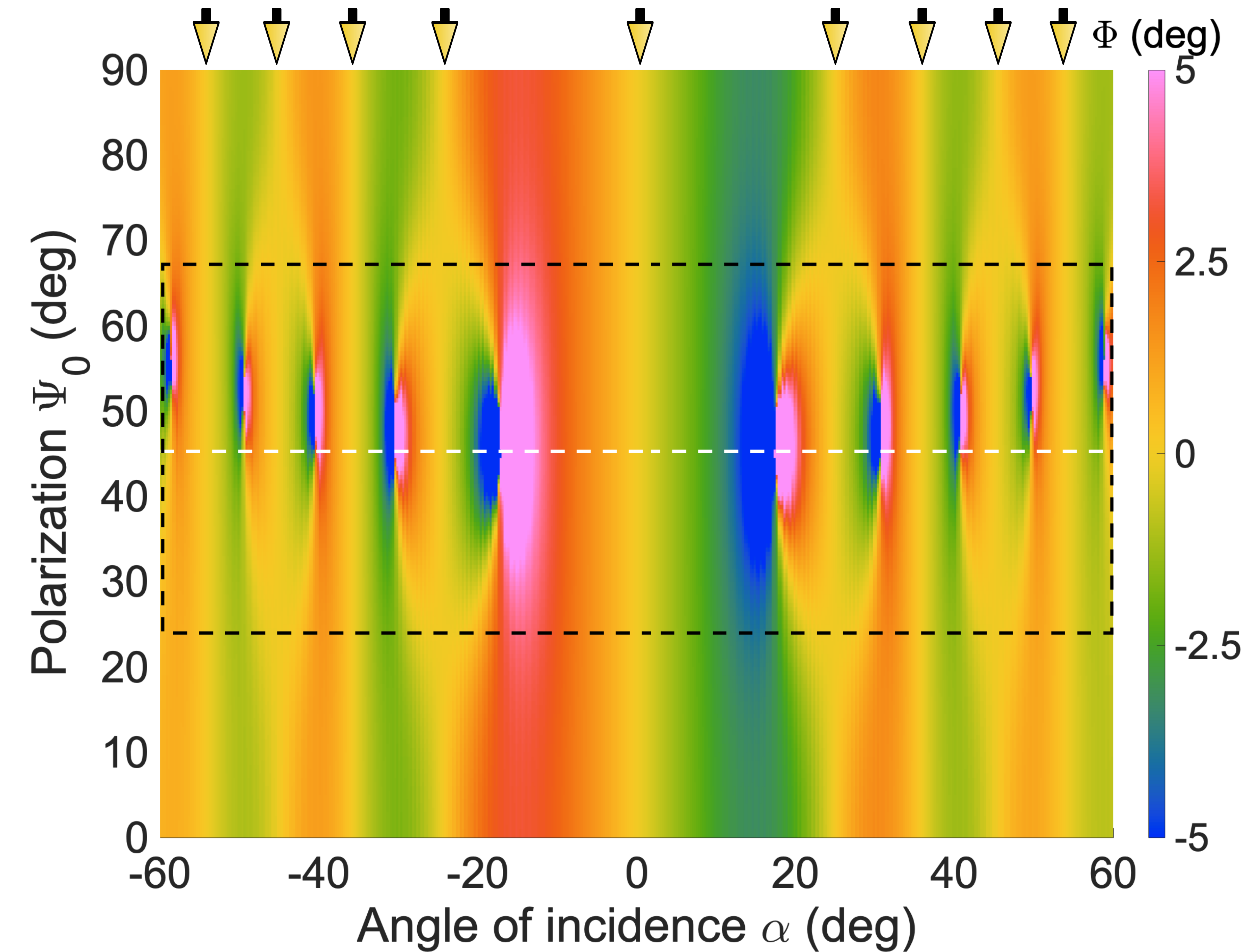} \\
(b) \\
\includegraphics[width=0.81\linewidth]{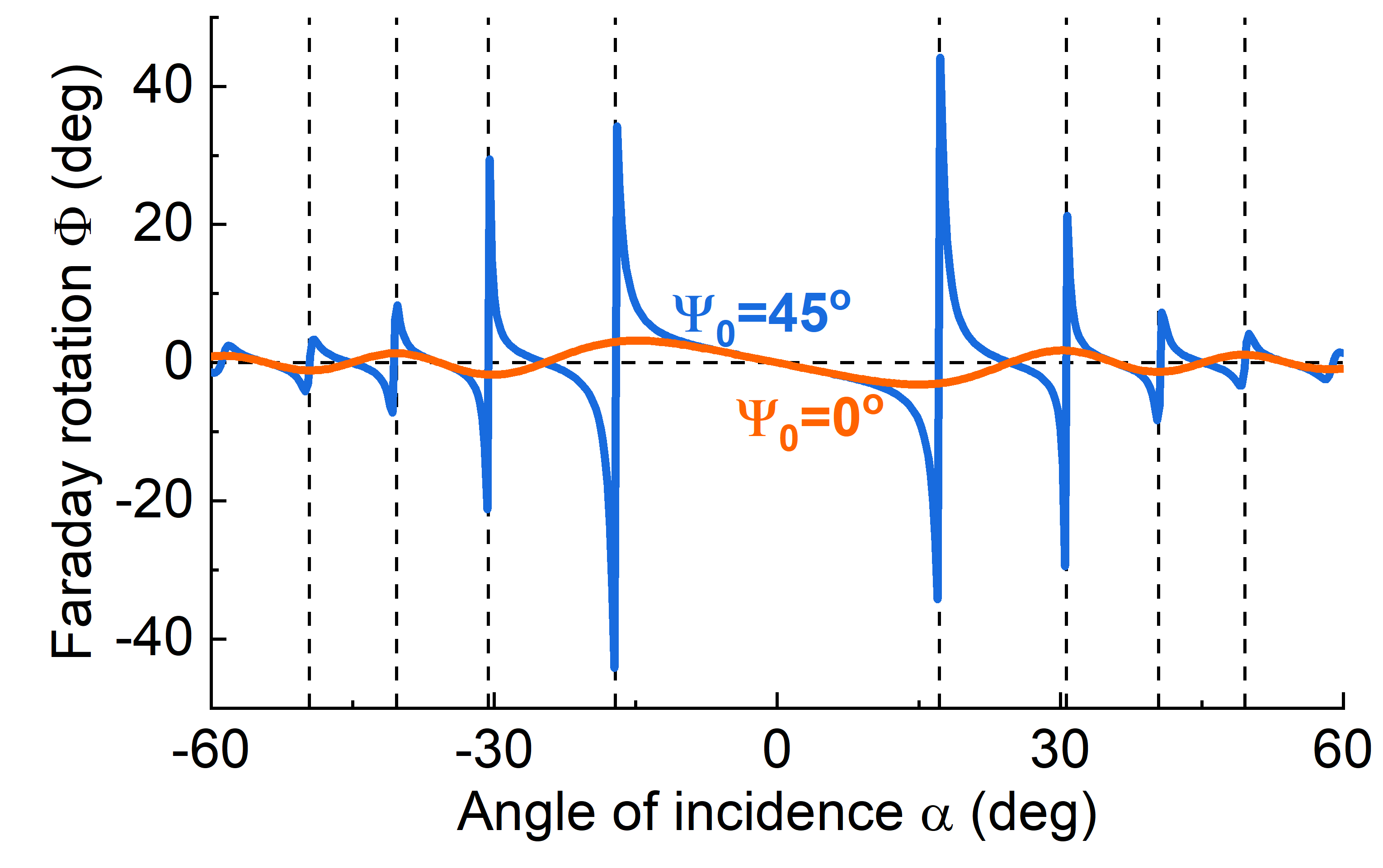}
\caption{\textcolor{teal}{Dependencies of the Faraday rotation on the angle of light incidence and its polarization $\Psi_0$ numerically obtained for the light with $\lambda=0.52\mu m$ wavelength propagating through a $\mathrm{FeBO_3}$ crystal of $d=160\mu$m thickness. (a) False-color plot with bright pink and blue areas corresponding to $|\Phi|>5^\circ$. Arrows mark angles where the birefringence phase shift $\phi=m\pi/2$. Black rectangle shows the region where the additional Faraday rotation sign change is observed comparing to the pure ordinary (or extraordinary) polarizations. (b) Angular dependence of the Faraday rotation for different initial light polarizations $\Psi_0$. Vertical dashed lines show the additional Faraday rotation sign changes observed for $\Psi_0=45^\circ$ polarization.}}
\label{fig: true_F}
\end{figure}

The peculiarity of the magneto-optical polarization effects in the birefringent crystals that to the best of our knowledge was not discussed before is that, according to Eq.~(\ref{Far_rotation}), the Faraday rotation strongly depends \textcolor{teal}{both on the initial polarization angle $\Psi_0$ and angle of incidence $\alpha$ (see Fig.~\ref{fig: true_F}). Similar to the case of the pure extraordinary polarization ($\Psi_0=0^\circ$) (Fig.~\ref{fig: MCD and Far}), the Faraday rotation experiences a sign-change even for an arbitrary initial polarization if the birefringence phase shift $\phi = m\pi/2, m\in \mathbb{Z}$ (the corresponding angles of incidence are marked with arrows in Fig.~\ref{fig: true_F}a). Besides, for the initial light polarization in the range $22.5^\circ\le\Psi_0\le67.5^\circ$ (marked as a black rectangle in Fig.~\ref{fig: true_F}a)), the additional Faraday rotation sign change occurs for $\sin^2\phi \sin^2 2\Psi_0 = \frac{1}{2}$. It is clearly seen in Fig.~\ref{fig: true_F}b (see dashed lines for the angles of incidence of $\alpha=\pm17^\circ, \pm30^\circ, \pm41^\circ, \pm50^\circ, \pm58^\circ$). Moreover, if the initial polarization $\Psi_0=45^\circ$ and the birefringent phase shift $\phi= (2m+1)\pi/4$, the denominator in Eq.~(\ref{Far_rotation}) equals to zero so that linear approximation used for its derivation is not valid anymore. Numerical simulations prove an enormous increase of the Faraday rotation in the vicinity of these conditions. Indeed, an unexpected phenomenon is that the Faraday rotation in the anisotropic film, if observed for these polarization and incidence angles, is much higher than in the isotropic medium with similar parameters (see Fig. S3 in Supplementary materials~\cite{suppl}).}  

\textcolor{teal}{Both Faraday rotation and purely optical birefringence affect light polarization modification during its propagation inside a magnetic crystal. Several experimental configurations allow to distinguish them and to measure pure magneto-optical Faraday rotation. For example, these are the schemes based on the crossed polarizers and measurements of the magnetic-field-induced changes of the polarizers angle corresponding to the minimum of the transmitted light intensity. Another approach is based on the light polarization modulation techniques via the photo-elastic modulator and measurement of the magnitude of corresponding temporal harmonics.}

\textcolor{teal}{At the same time, in most of the practically interesting devices where the Faraday rotation is used, such as magneto-optical microscopes~\cite{mccord2015progress}, magneto-optical modulators and isolators, the pump-probe setups with balanced detection schemes, the cumulative optical and magneto-optical effects are measured rather than pure magneto-optical ones. Therefore, it is important to analyze how these cumulative effects reveal themselves in detail. Most of these devices are based on an additional polarizer oriented at $45^\circ$ angle to the incident polarization. Magneto-optical rotation is defined as the relative variation of the magnetic crystal transmittance after light passes through the polarizer and thus acquires both the purely optical and the magneto-optical contributions to the polarization rotation. Indeed, the intensity of light transmitted through the magnetized sample and polarizer $I^{45^\circ}_{\pm M}$: 
\begin{equation}
    I^{45^\circ}_{M}=T_M\cdot \frac{|\cos({45^\circ}+\Delta\Psi_M)|^2}{|\cos({45^\circ}+\Delta\Psi_M)|^2+|\sin({45^\circ}+\Delta\Psi_M)|^2},
    \label{I_pi.4}
\end{equation}
depends on the transmittance $T_M$ and $\Delta\Psi_M$, a total variation of complex polarization angle that includes contributions of the optical and magneto-optical ellipticities along with purely optical and Faraday polarization rotations. The quantity which is actually measured during such kind of the magneto-optical measurements,
\begin{equation}
    \Delta = \frac{I^{45^\circ}_{+M}-I^{45^\circ}_{-M}}{I^{45^\circ}_{+M}+I^{45^\circ}_{-M}}\times 100\%,
\end{equation}
thus has the contributions of the whole set of optical and magneto-optical effects~\eqref{I_pi.4}. It corresponds to the pure Faraday rotation angle $\Phi= \frac{1}{2}\arcsin \Delta$ only for a particular case of the linearly polarized light outcoming from the sample.} The physical meaning of $\Delta$ \textcolor{teal}{in general case} is the optical contrast between the +M and -M states  \textcolor{teal}{observed with the additional polarizer. Higher $\Delta$ values correspond to a better resolution of the magneto-optical microscope, a larger magneto-optical light modulation efficiency, or higher sensitivity of the probe to the spin dynamics, depending on the setup application. Therefore, $\Delta$ is an important characteristic of a magneto-optical system.}

\begin{figure}[htbp]
(a)\\
\includegraphics[width=0.94\linewidth]{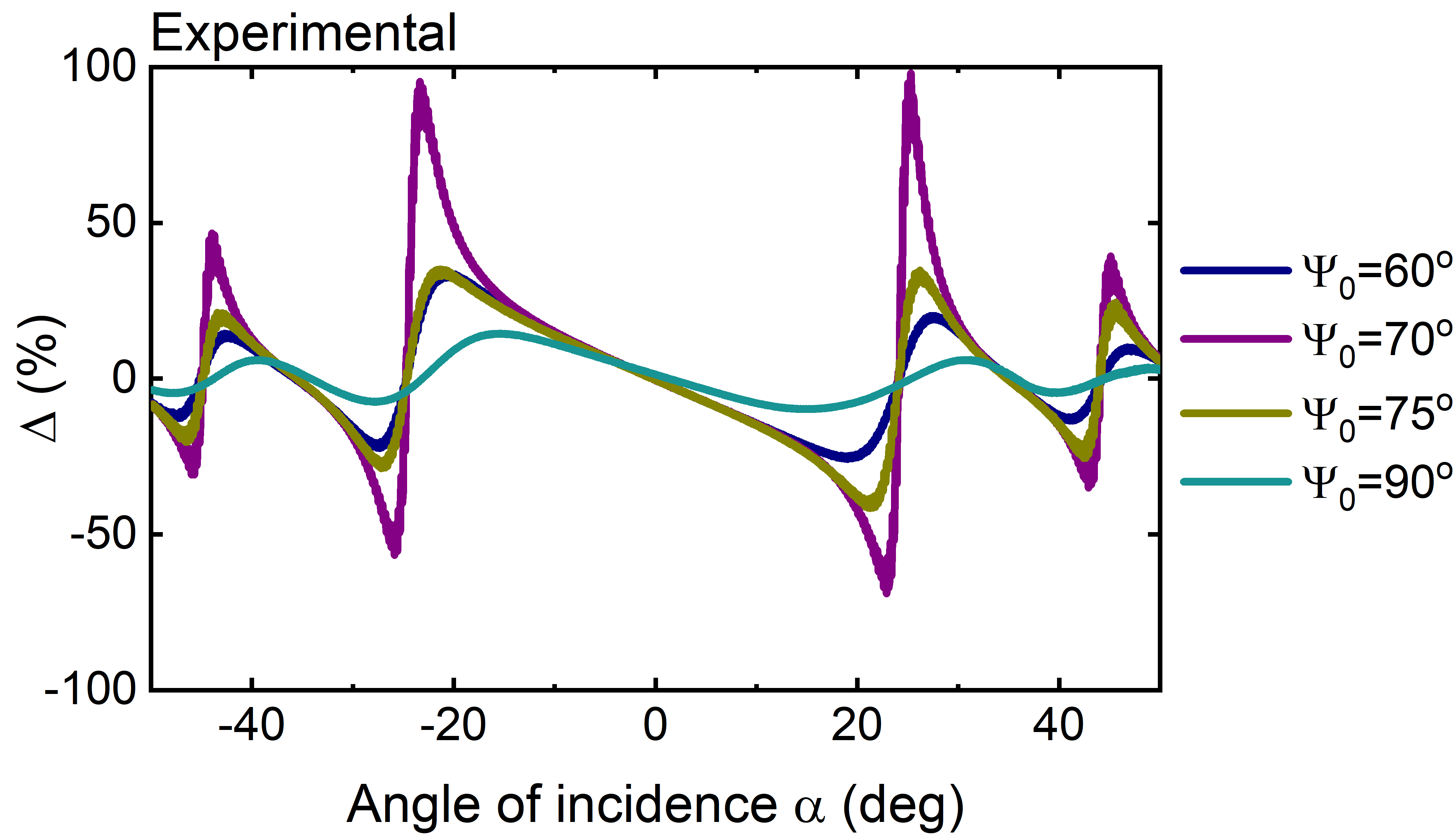}\\
\centering
(b)\\
\includegraphics[width=0.99\linewidth]{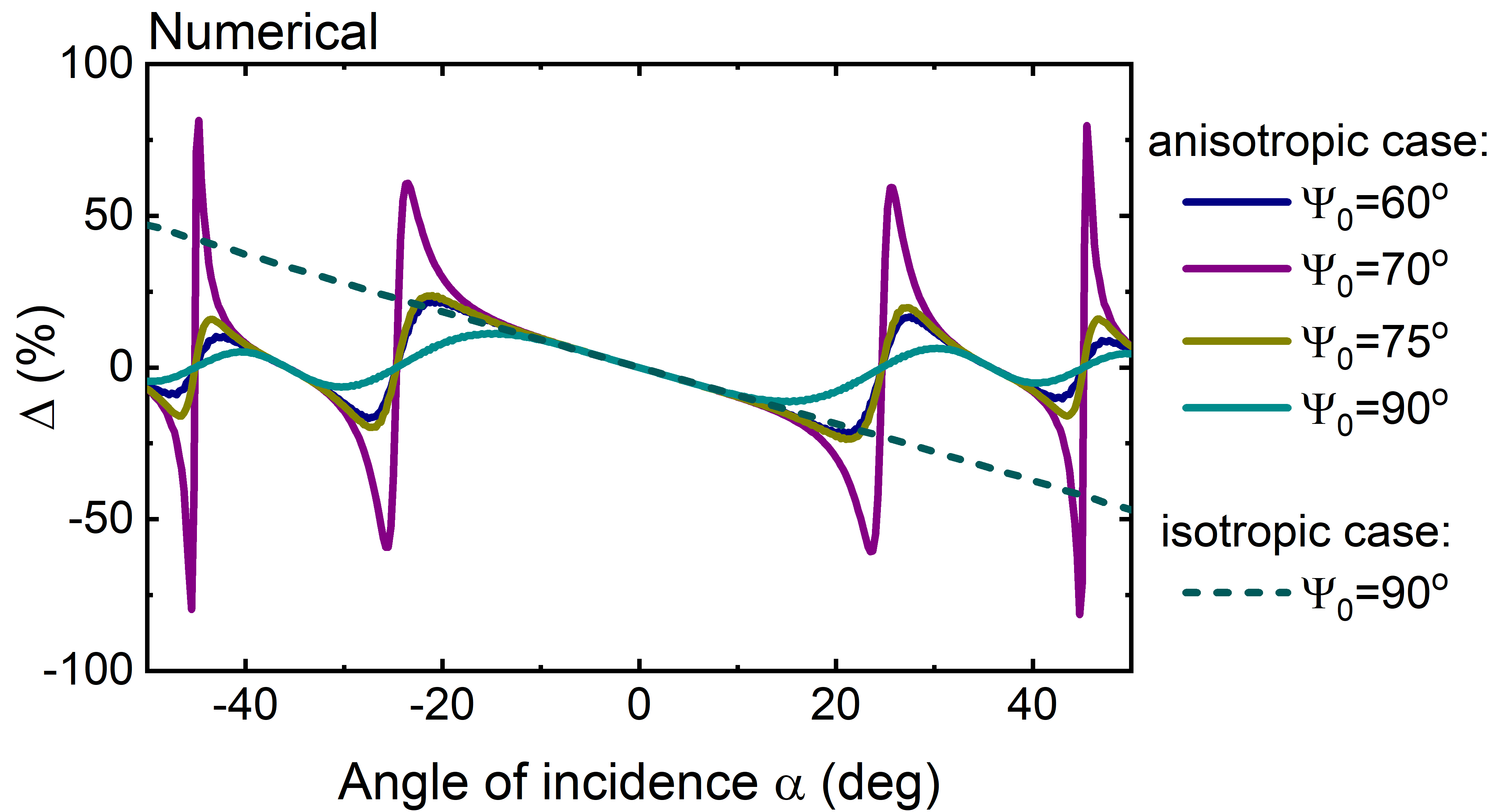}
\caption{(a) Measured and (b) simulated magneto-optical activity $\Delta$ for different polarization angles $\Psi_0$ of the linearly polarized incident light (see the legend).}
\label{fig: real delta}
\end{figure}

Fig.~\ref{fig: real delta} shows the experimental and numerical results for the considered crystal (see also Fig. S3 in Supplementary materials~\cite{suppl}). For the \textcolor{teal}{pure ordinary $\Psi_0=90^\circ$ or extraordinary $\Psi_0=0^\circ$} light polarization, $\Delta < 10\%$ which corresponds to the Faraday rotation of less than $3^\circ$. Simulation results for the isotropic case predict the linear growth of $\Delta$ depending on the incidence angle with the maximum value of $\Delta=50\%$ (dashed line on Fig.~\ref{fig: real delta}b). However, for the polarization of light close to $\Psi_0=70^\circ$, a very strong magneto-optical modulation is observed reaching $\Delta \approx 100\%$ due to the anisotropic properties of the real material. The difference between \textcolor{teal}{the initial polarization $\Psi_0$} optimal for the enhancement of the pure Faraday rotation $\Phi$ (Fig.~\ref{fig: true_F}) and $\Delta$ (Fig.~\ref{fig: real delta}) is explained by \textcolor{teal}{the contribution of purely optical effects} to $\Delta$. Therefore, by tuning the \textcolor{teal}{light polarization and angle of incidence} it is possible to achieve almost 100$\%$ contrast between the +M and -M states.

The situation, when the crystal optical birefringence results in a drastic increase of the observed magneto-optical effects, is quite unexpected and certainly should be exploited for the magneto-optical measurements. Moreover, the sharp $\Delta$ peaks of different signs are very close to each other in the angular spectra, so that the derivative $\partial \Delta / \partial \alpha$ enormously grows near these angles (see, for example, $\alpha\sim 15^\circ$ in Fig.~\ref{fig: real delta}). As $\mathrm{FeBO_3}$ crystal is known to be very sensitive to mechanical stresses~\cite{khizhnyi2017acoustic}, such dependence could be used for various magneto-acousto-optical devices where the acoustic pulse controls the birefringence $\Delta \varepsilon$, and the magnetic field is responsible for the light modulation.

To sum up, we have studied experimentally, analytically, and numerically the peculiarities of the magneto-optical effects in the $\mathrm{FeBO_3}$ crystal characterized by large optical and magnetic anisotropies. We have shown that optical birefringence might lead to a significant increase of the observed magneto-optical effects. The key to making benefits from the birefringence is to adjust the light polarization and incidence angle properly. Though the phenomenon of the birefringence mediated enhancement of the magneto-optical effects has been demonstrated here on the $\mathrm{FeBO_3}$ crystals, it is quite general and can be found in other types of optically anisotropic magnetic crystals. The most \textcolor{teal}{intriguing possible} practical application of the observed phenomenon might be the ultrafast light modulation with up to 100$\%$ efficiency and THz frequency, which is impossible both in isotropic ferromagnetic materials and anisotropic antiferromagnet crystals without polarization tuning. Other applications of the described effect are expected to include imaging of the antiferromagnets, sensing, and magnetometry. 

\medskip
This work was financially supported by the Ministry of Science and Higher Education of the Russian Federation, Megagrant project No. 075-15-2019-1934.

The authors thank Prof. A.K. Zvezdin and Dr. A.N. Kalish for fruitful discussions.

\medskip

\noindent\textbf{Disclosures.} The authors declare no conflicts of interest.

\nocite{*}

\providecommand{\noopsort}[1]{}\providecommand{\singleletter}[1]{#1}%

\end{document}